\documentclass{elsart}
\usepackage{times}
\usepackage{epsfig}
\usepackage{marvosym}
\usepackage{amsmath}
\usepackage{amsfonts}
\usepackage{amssymb}
\begin{document}
\bibliographystyle{roman}

\journal{Nuclear Instruments and Methods A}

\def\hb{\hfill\break}
\def\MeV{\rm MeV}
\def\GeV{\rm GeV}
\def\TeV{\rm TeV}

\def\m{\rm m}
\def\cm{\rm cm}
\def\mm{\rm mm}
\def\lam{$\lambda_{\rm int}$}
\def\rad{$X_0$}
 
\def\NIM{Nucl. Instr. and Meth.~}
\def\ieee {{IEEE Trans. Nucl. Sci.~}}

\def\etal{{\it et al.}}
\def\eg{{\it e.g.,~}}
\def\ie{{\it i.e.,~}}
\def\cf{{\it cf.~}}
\def\etc{{\it etc.~}}
\def\vs{{\it vs.~}}
\begin{frontmatter}
\title{Contributions of \v{C}erenkov Light to the Signals from Lead Tungstate Crystals}

\author{N. Akchurin$^a$, L. Berntzon$^a$, A. Cardini$^b$, R. Ferrari$^c$, G. Gaudio$^c$,}
\author{J. Hauptman$^d$, H. Kim$^a$, L. La Rotonda$^e$, M. Livan$^c$, E. Meoni$^e$,}
\author{H. Paar$^f$, A. Penzo$^g$, D. Pinci$^h$, A. Policicchio$^e$,} 
\author{S. Popescu$^{i,}$\thanksref{Leave}}
\author{G.~Susinno$^e$, Y. Roh$^a$, W. Vandelli$^c$ and R. Wigmans$^{a,}$\thanksref{Corres}}

\address{$^a$ Texas Tech University, Lubbock (TX), USA\\
$^b$ Dipartimento di Fisica, Universit\`a di Cagliari and INFN Sezione di Cagliari, Italy\\
$^c$ Dipartimento di Fisica Nucleare e Teorica, Universit\`a di Pavia and INFN Sezione di Pavia, Italy\\
$^d$ Iowa State University, Ames (IA), USA\\
$^e$ Dipartimento di Fisica, Universit\'a della Calabria and INFN Cosenza, Italy\\
$^f$ University of California at San Diego, La Jolla (CA), USA\\
$^g$ INFN Trieste, Italy\\
$^h$ Dipartimento di Fisica, Universit\`a di Roma ''La Sapienza''  and INFN Sezione di Roma\\
$^i$ CERN, Gen\`eve, Switzerland}
\thanks[Leave]{On leave from IFIN-HH, Bucharest, Romania.}
\thanks[Corres]{Corresponding author.
              Email wigmans@ttu.edu, fax (+1) 806 742-1182.}

\begin{abstract}
Results are presented of detailed measurements of the signals generated by high-energy electrons
and muons in lead tungstate crystals. A significant fraction of the light produced in these crystals and detected by photomultiplier tubes is the result of the \v{C}erenkov mechanism. This is concluded from the angular dependence of the signals and from their time structure.
Depending on the orientation of the crystals and on the particle type, \v{C}erenkov light may account for up to 15\% of the total signals. 

\end{abstract}

\end{frontmatter}

\section{Introduction}

In recent years, there has been a growing interest in lead tungstate (PbWO$_4$) crystals as detectors
for high-energy particles. At CERN's Large Hadron Collider, both the CMS \cite{CMS} and ALICE \cite{ALICE} experiments
are completing very large electromagnetic (em) calorimeter systems consisting of these crystals. Smaller
detectors of this type are either operating or planned, for example for the PANDA \cite{Panda} and HYCAL \cite{HYCAL} experiments. Lead tungstate crystals are attractive as detectors for em showers because of their high
density, which implies a short radiation length and Moli\`ere radius, their fast signals and their relative insensitivity to the effects of radiation damage. On the downside, we mention the small light yield, less than 1/300 of the light yield of the widely used NaI(Tl) and CsI(Tl) crystals \cite{Zhu}.
Because of this small light yield, and the large effective $Z$ value, it is reasonable to assume that a significant fraction of the light produced by PbWO$_4$ crystals is actually the result of \v{C}erenkov radiation, rather than molecular de-excitation.
 
We have shown previously that the combined availability of \v{C}erenkov and scintillation signals for hadronic showers makes it possible to eliminate the effects of the dominating source of fluctuations
in calorimetric hadron detection, and thus considerably improve hadronic calorimeter 
performance \cite{DREAMhad}.  However, because of the small \v{C}erenkov light yield, sampling
calorimeters are less than ideal for taking full advantage of this. On the other hand, homogeneous calorimeters whose light signals could be split into scintillation and \v{C}erenkov components hold great promise for high-quality hadron calorimetry \cite{elba}.

For these reasons, we set out to measure the composition of the signals produced by PbWO$_4$ crystals\footnote{The PbWO$_4$ crystals used for these studies were provided by the ALICE Collaboration, who use them for their PHOS calorimeter.}. 
In Sections 2 and 3, we describe the methods used to determine the \v{C}erenkov component of the measured signals, and the
experimental setup in which the crystals were tested. In Section 4, we discuss the experimental
data that were taken and the methods used to analyze these data. 
In Section 5, the experimental
results are presented and discussed. A summary and conclusions are presented in Section 6.    

\section{Methods to distinguish \v{C}erenkov from scintillation light.}

If one wants to distinguish the contributions from the \v{C}erenkov and scintillation components to the
signals from crystals that generate a mixture of these, such as PbWO$_4$, one could use one or several
of the following characteristics: 
\begin{enumerate}
\item {\sl Directionality}. The \v{C}erenkov light is emitted at a fixed angle with respect to the momentum vector of the particle that generates it, while the scintillation light is isotropically emitted.
\item {\sl Time structure}. The \v{C}erenkov light is {\sl prompt}, whereas scintillation processes have one or several characteristic decay times.
\item {\sl The spectrum}. The \v{C}erenkov light is emitted with a characteristic $\lambda^{-2}$ spectrum, while the scintillation processes have their own characteristic spectra.
\item {\sl Polarization}. Contrary to scintillation light, \v{C}erenkov light is polarized. 
\end{enumerate}

In the present study, we have exploited the first two characteristics.
It is a well known fact that the light yield of PbWO$_4$ crystals is very temperature dependent. It changes by -2.3\%  per degree Celsius.
This is obviously only true for the {\sl scintillation} light. Therefore, the fraction of the light that is produced
by the \v{C}erenkov mechanism is expected to increase with temperature. Our  measurements have been performed at room temperature. One should keep this in mind when translating our results to the ALICE experiment, from which the crystals we tested were borrowed. The ALICE PbWO$_4$ calorimeter operates at a much lower temperature ($-15^\circ$C), where the relative contributions of \v{C}erenkov light to the signals are correspondingly smaller\footnote{The PbWO$_4$ ECAL of the CMS experiment is designed to operate at 18$^\circ$C.}.

\subsection{Directionality}

\v{C}erenkov light is emitted by charged particles traveling faster than $c/n$, the speed of light in the medium with refractive index $n$ in which this process takes place. The light is emitted at a characteristic angle, $\theta_C$, defined by $\cos \theta_C = 1/\beta n$.
In the case of sufficienty relativistic particles (\ie $\beta \sim 1$) traversing PbWO$_4$ crystals ($n = 2.2$), $\theta_C \sim 63^\circ$.
\begin{figure}[htb]
\epsfysize=7cm
\centerline{\epsffile{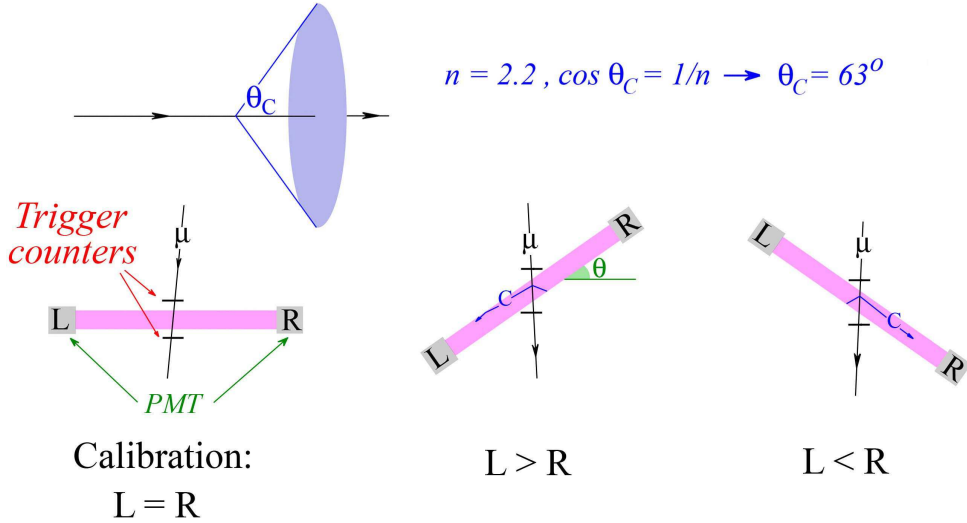}}
\caption{\small
Principle of the asymmetry measurement used to establish the contribution of \v{C}erenkov light to the 
signals from the PbWO$_4$ crystals. Depending on the orientation, this directionally emitted light contributes differently to the signals from the left and right photomultiplier tubes. }
\label{LRprinciple}
\end{figure}

In order to detect the contribution of \v{C}erenkov light to the signals from a PbWO$_4$ crystal, we
equipped both ends of the crystal with a photomultiplier tube (PMT). By varying the detector {\sl orientation} with respect to the direction of the incoming particles, 
a contribution of \v{C}erenkov light would then manifest itself as an angle-dependent asymmetry. 
This is illustrated in Figure \ref{LRprinciple}, which shows the setup of the initial measurements we performed with a cosmic-ray telescope to test this principle \cite{elba}. The PMT gains were equalized
for the leftmost geometry, in which the crystal was oriented horizontally. 
By tilting the crystal through an angle ($\theta$) such that the axis of the crystal is at the \v{C}erenkov angle 
$\theta_C$ with respect to the particle direction, \v{C}erenkov light produced by the cosmic rays traversing the trigger counters would be preferably detected in either the $L$ (central geometry) or $R$ (rightmost geometry)
PMT. By measuring the response asymmetry $(R-L)/R+L)$ as a function of the tilt angle $\theta$, the contribution of \v{C}erenkov light to the detector signals could be determined.

The initial cosmic-ray measurements indicated that the contribution of \v{C}erenkov light was at the level of 15 - 20\% \cite{elba}. Because of the extremely low event rates and the tiny signals (typically 20 - 30 MeV), we decided to perform systematic studies using particle beams. The results of these studies are the topic of the present paper.

\subsection{Time structure}

The scintillation process in PbWO$_4$ has a decay constant of $\sim 10$ ns, whereas the \v{C}erenkov component of the signals is prompt. In the cosmic-ray measurements mentioned above, we also studied the time structure of the signals, using a fast oscilloscope capable of storing both pulse shapes simultaneously. 
\begin{figure}[htb]
\epsfysize=6.5cm
\centerline{\epsffile{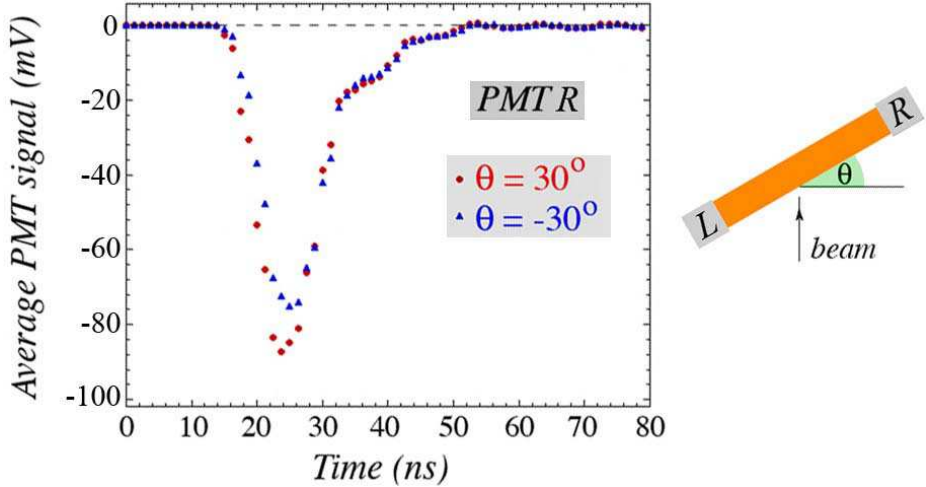}}
\caption{\small
Average time structure of the signals measured with the PMT reading out one end ($R$) of a PbWO$_4$ crystal traversed by 10 GeV electrons, for two different orientations of the crystal. At  $\theta = 30^\circ$, \v{C}erenkov light contributes to the signals, at $\theta = - 30^\circ$, it does not.}
\label{tstruc}
\end{figure}
These measurements qualitatively confirmed the expected differences between the signals measured in the $R$ and $L$ PMTs \cite{elba}. 
However, low statistics and very small signals limited the quality of the information that could be derived from these measurements.
In the beam measurements described here, we measured the signal shapes with very fast Flash ADCs (effective sampling frequency 800 MHz). As is shown in Figure \ref{tstruc}, this turned out to be a wonderful experimental tool. In a period of a few days, detailed pulse shapes were recorded for millions of events. The figure clearly shows the additional prompt signal component that appears when the crystal is rotated from a position in which \v{C}erenkov light does not contribute to the signals ($\theta = -30^\circ$) to a position where it does ($\theta = 30^\circ$). The trailing edge of the PMT signals is not affected by this rotation and is indeed in great detail (including the effects of reflections in the signal cables) identical for these two pulse shapes.

\section{Experimental setup}

\subsection{Detector and beam line}

The measurements described in this paper were performed in the H4 beam line of the Super Proton Synchrotron at CERN. Our detector was a PbWO$_4$ crystal with a length of 18 cm and a cross section of $2.2\times 2.2$ cm$^2$. The transverse dimension, relevant for our measurements, corresponds to
2.5 radiation lengths.  The light produced by particles traversing this crystal was read out by two photomultiplier tubes\footnote{Hamamatsu R5900U, 10-stage, bialkali photocathode, borosilicate window.}, located at opposite ends. 
In order to reduce the light trapping effects of the large refractive index of PbWO$_4$, the PMTs were coupled to the crystal by means of silicone ``cookies'' ($n = 1.403$).
\begin{figure}[htb]
\epsfysize=7cm
\centerline{\epsffile{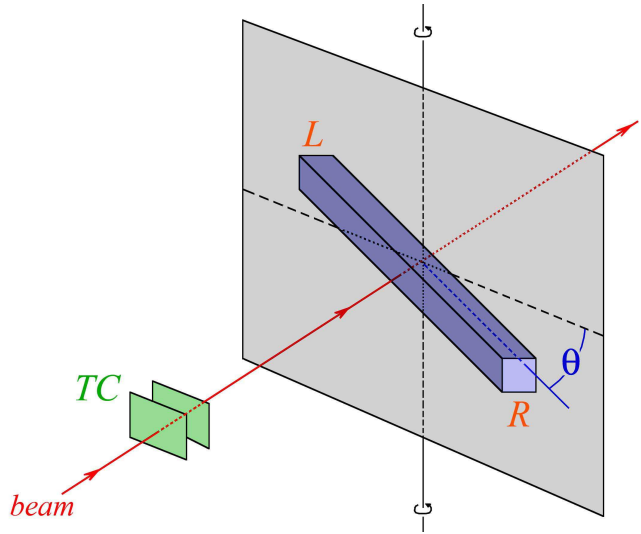}}
\caption{\small
Experimental setup in which the beam tests were performed.}
\label{beamtest}
\end{figure}

This crystal
was mounted on a platform that could rotate around a vertical axis. The crystal was oriented in the horizontal plane and the rotation axis went through its geometrical center. The particle beam was
also steered through this center, as illustrated in Figure \ref{beamtest}. The angle $\theta$, which is frequently used in the following, represents the angle between the crystal axis and a plane perpendicular to the beam line.
The angle increases when the crystal is rotated such that the crystal axis L-R approaches the direction of the 
traveling beam particles. The crystal orientations shown in Figures \ref{tstruc} and
\ref{beamtest} correspond thus to  $\theta > 0$ and $\theta < 0$, respectively.

Two small scintillation counters provided the signals that were used to trigger the data acquisition system.
These Trigger Counters (TC) were 2.5 mm thick, and the area of overlap was 6$\times$6 cm$^2$. A coincidence between the logic
signals from these counters provided the trigger.

\subsection{Data acquisition}

Measurement of the time structure of the crystal signals formed a very important part of the tests
described here. In order to limit distortion of this structure as much as possible, we used 15 mm thick air-core cables to transport the detector signals to the counting room. Such cables were also used for
the signals from the trigger counters, and these were routed such as to
minimize delays in the DAQ system\footnote{We measured the signal speed to be 0.78$c$ in these cables}.

Depending on the desired type of information, the crystal signals were either sent to a charge ADC, or to
the FADC. The response asymmetry measurements were based on the digitized integrated charge, the time structure was measured with the Flash ADC\footnote{Dr. Struck SIS3320,  http://www.struck.de/sis3320.htm}, which digitized the amplitude of the signals at a rate of 200 MHz. During a time interval of 80 ns, 16 measurements of the amplitude were thus obtained. In order to further increase this rate, and thus improve the time resolution of this measurement, we used several input channels for each signal.
The crystal signals  were split (passively, with correct impedance matching) into 4 equal parts at the counting room end.
These 4 signals were measured separately in 4 different channels of the FADC module. Signals 2, 3 and 
4 were delayed by 1.25 ns, 2.50 ns and 3.75 ns with respect to signal 1. By using the FADC module in this way, the time structure of the signals was thus effectively measured with a resolution of 1.25 ns (800 MHz).

The quality of the information obtained in this way is illustrated in Figure \ref{tstruc}, which shows the 
average time structure of the signals from 10 GeV electrons traversing the crystal for $\theta = 30^\circ$ and $-30^\circ$, respectively.

The charge measurements were performed with 12-bit LeCroy 1182 ADCs.
These had a sensitivity of 50 fC/count and a conversion time of 16 $\mu$s.
The ADC gate width was 100 ns, and the calorimeter signals arrived $\sim 20$ ns after the start of the gate.

The data acquisition system used VME electronics.
A single VME crate hosted all the needed readout and control boards.
The trigger logic was implemented through NIM modules and the signals were sent 
to a VME I/O register, which also collected the spill and the global 
busy information. The VME crate was linked to a Linux based computer 
through an SBS 620\footnote{http://www.gefanucembedded.com/products/457} 
optical VME-PCI interface that allowed memory 
mapping of the VME resources via an open source driver\footnote{http://www.awa.tohoku.ac.jp/$\sim$sanshiro/kinoko-e/vmedrv/}. The computer was equipped with a 2 GHz Pentium-4 
CPU, 1 GB of RAM, and was running a CERN SLC 4.3 operating system\footnote{http://linux.web.cern.ch/linux/scientific4/}.

The data acquisition was based on a single-event polling mechanism and 
performed by a pair of independent programs that communicated
through a first-in-first-out buffer, built on top of a 32 MB shared 
memory. Only exclusive accesses were allowed and concurrent requests were 
synchronised with semaphores. The chosen scheme 
optimized the CPU utilization and increased the data taking efficiency by 
exploiting the bunch structure of the SPS, where beam particles were provided to
our experiment during a spill of 4.8 s, out of a total cycle time of 16.8 s.
During the spill, the readout program collected data from the VME modules and 
stored them into the shared memory, with small access times. During the remainder of the SPS cycle, a 
recorder program dumped the events to the disk. Moreover, the buffer
presence allowed low-priority monitoring programs to run (off-spill) in 
spy mode. With this scheme, we were able to reach a data acquisition rate 
as high as 2 kHz, limited by the FADC readout time. 
The typical event size was $\sim 1$ kB.  
All detector signals were monitored on-line.

\subsection{Calibration of the detectors}

The absolute calibration of the signals generated by the crystal was not a major concern in these tests.
On the other hand, it was absolutely essential that the gains of the 2 PMTs, $L$ and $R$, that 
collected the light generated in the crystals at the two opposite ends of the crystal were equalized.
We used 10 GeV electrons for that purpose.
The crystal was oriented such that the beam entered the detector perpendicular to the 
crystal axis ($\theta = 0$), so that any \v{C}erenkov light generated by the beam particles would be
observed in the same proportion by both PMTs.
The high voltages were chosen such that the average signals were about 300 ADC counts above the pedestal value. Off-line, the calibration constants were fine-tuned such as to equalize the responses of the two PMTs.

\section{Experimental data}

The crystals were exposed to beams of 150 GeV $\mu^+$ and 10 GeV electrons.
The angle $\theta$ between the crystal axis and the plane perpendicular to the beam line was varied
from $-45^\circ$ to $45^\circ$, in steps of $7.5^\circ$. At each angle, 100 000 events were collected
for the response asymmetry measurements, and another 100 000 for the time structure.

Since the particles traversed the detector perpendicular to the longitudinal crystal axis, the effective
thickness of the crystal was only a few radiation lengths ($2.5 X_0/\cos \theta$) in this setup.
In order to probe the em showers at greater depth, we also performed a series of measurements in
which the electrons traversed 4 cm of lead ($\sim 7 X_0$) installed directly upstream of the crystal. In this way, the light generated in the crystals reflected the particle distribution just beyond the shower maximum, at a depth of $7-10 X_0$. To avoid introducing too large a change in this effective depth, the latter measurements were limited to angles $\theta$ ranging from $-30^\circ$ to $30^\circ$. 
Separate measurements were performed of the response asymmetry and of the time structure. 
As before, 100 000 events were collected for each run.

\section{Experimental results}

\subsection{Left-Right asymmetry}

We define the response asymmetry as the ratio $(R - L)/(R + L)$, where $R$ and $L$ represent the
average signals measured in the PMTs $R$ and $L$ for the same events. Since these signals were equalized for $\theta = 0$, any non-zero value in this ratio is indicative for a non-isotropic component
in the light generated in the crystals, \ie \v{C}erenkov light. 

The relationship between this response asymmetry (to be called $\alpha$ in the following) and the relative contribution of \v{C}erenkov light to the PMT signals\footnote{It should be emphasized that this discussion concerns the PMT signals, and {\sl not} the numbers of photons produced by the different mechanisms. For the latter, differences in production spectra and photocathode quantum efficiencies would have to be taken into account.} can be seen as follows.
If we call the relative contributions of \v{C}erenkov light to the $R$ and $L$ signals $\epsilon_R$ and $\epsilon_L$,
respectively (with  $\epsilon_R$ and  $\epsilon_L$ normalized to the scintillator signals in each channel), then 
\begin{equation}
\alpha ~= ~{{\epsilon_R -  \epsilon_L}\over{2 +  \epsilon_R +  \epsilon_L}}
\label{eq1}
\end{equation}
This ratio reaches its maximum possible value when \v{C}erenkov light reaches only one of the PMTs, \eg $R$.
In that case,  $\epsilon_L = 0$,  $\alpha =  \epsilon_R/(2 +  \epsilon_R)$, and the relative contribution
of \v{C}erenkov light to the {\sl total} signal from this PMT  equals
\begin{equation}
f_C ~=~ {\epsilon_R\over{1 +  \epsilon_R}} ~=~ {2\alpha\over{1 + \alpha}}
\label{eq2}
\end{equation} 
This situation may occur when a single relativistic charged particle traverses the crystal. Depending on the orientation of the crystal and the index of refraction, the acceptance of the
\v{C}erenkov light emitted by that particle may in that case be limited to one PMT only. 
Equation \ref{eq1} shows that when the \v{C}erenkov light produced in the crystal is shared between both PMTs,
\ie when both  $\epsilon_R$ and $\epsilon_L$ are non-zero, then the measured value of the asymmetry is smaller
than the maximum possible value mentioned above, and Equation \ref{eq2} underestimates the contribution of \v{C}erenkov light to the signals. As we shall see below, this situation occurs in developing showers.
%
\begin{figure}[htb]
\epsfysize=9cm
\centerline{\epsffile{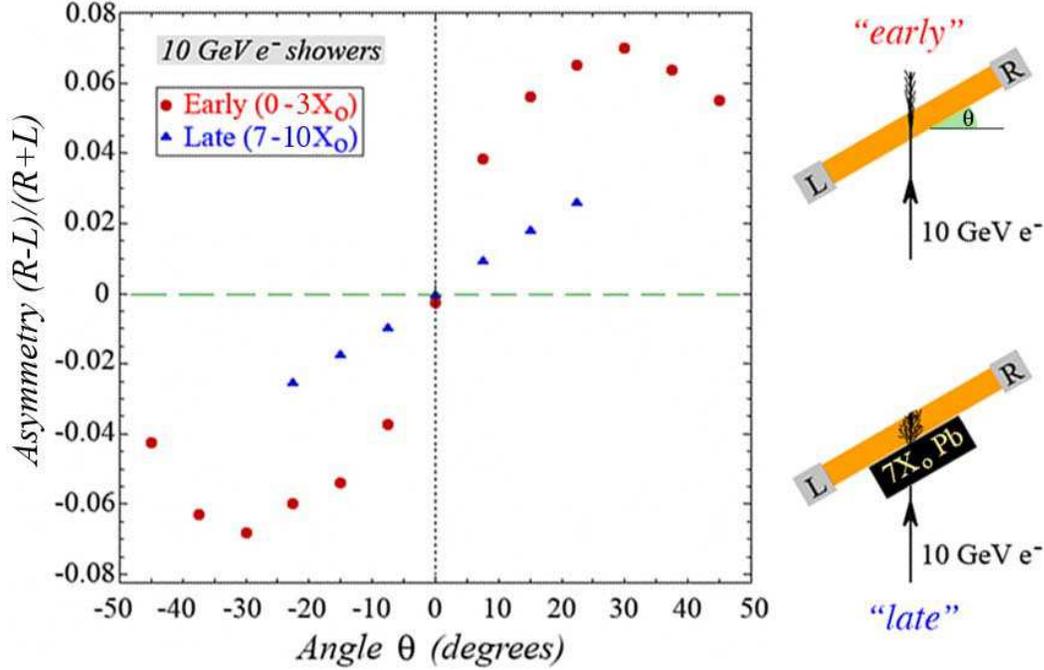}}
\caption{\small
Left-right response asymmetry measured for 10 GeV electrons showering in a $2.5 X_0$ thick 
PbWO$_4$ crystal, as a function of the orientation of the crystal (the angle $\theta$). Results are shown for the early and the late components of the showers. The latter measurements were obtained by placing 4 cm of lead upstream of the crystal. }
\label{Asymel2}
\end{figure}

Figure \ref{Asymel2} shows the response asymmetry measured for 10 GeV electrons, as a function of the angle $\theta$. This curve exhibits exactly the characteristics expected from a contribution of \v{C}erenkov light:
\begin{itemize}
\item It is symmetric around $\theta = 0$, \ie the measured response asymmetry at an angle $\theta$ is equal to that at $- \theta$. 
\item The asymmetry rapidly increases when the crystal is rotated, reaches a maximum value for
$\theta = \pm 30^\circ$, and declines again for larger angles. This reflects the changing acceptance of 
\v{C}erenkov light, as we verified with a very simple light tracing Monte Carlo simulation.
\item The maximum asymmetry is measured for an angle that is close to 
 the expected value of $90^\circ -\theta_C = 27^\circ$, at which angle the acceptance for \v{C}erenkov light
 is largest in this geometry. 
\end{itemize}

The maximum asymmetry amounts to 0.07, and we conclude from this that when the crystal is oriented at
$\theta = \pm30^\circ$, $\sim 13\%$ of the signal measured in the PMT that is optimally located for detecting \v{C}erenkov light, is indeed \v{C}erenkov light (Equation \ref{eq2}). As the orientation of the crystal is changed, this percentage decreases, reflecting the reduced acceptance for \v{C}erenkov light of the PMTs. At $\theta = 0$, the total acceptance for \v{C}erenkov light in both PMTs combined reaches its minimum value.

The above arguments are, strictly speaking, only valid for particles that traverse the crystal in a direction parallel to the beam. Upon traversing the crystal, the beam electrons lose a large fraction of their energy (typically $> 80\%$) radiating bremsstrahlung photons. The relativistic electrons and positrons produced when these photons convert in the crystal travel also predominantly in the same direction. For the purpose of this experiment, the early part of the electromagnetic showers probed in this measurement thus resembles a collection
of particles traveling all approximately in the same direction, \ie parallel to the beam line. 

However, as the shower develops, so does its isotropic component. This component is primarily due to shower electrons generated in Compton scattering or through the photoelectric effect. In fully contained em showers, this component is responsible for about half of the signal \cite{RWbook}. It is also thanks to this component that
$0^\circ$ quartz-fiber calorimeters, such as the CMS HF, produce meaningful signals \cite{RSI}.

For this reason, we also wanted to measure the effect of this increased isotropy on the left/right response asymmetry. By placing 4 cm of lead directly upstream of the crystal, the electron showers developed in a lead/PbWO$_4$ combination. For 10 GeV electrons, the shower maximum was located at a depth of $\sim 5 X_0$, (\ie inside the lead absorber), and the crystal probed the light produced at a depth of $7-10 X_0$.  

Figure \ref{Asymel2} also shows the response asymmetry measured for this light. The asymmetry is considerably smaller than for the light produced in the early part of the shower, by about a factor of three.
Yet, the characteristics of the $(R-L)/(R+L)$ curve indicate that also in this case, the asymmetry is the result of the contribution of \v{C}erenkov light to the signals. The reduction in the net directionality of the measured light indicates the importance of the isotropic shower component. 

In another paper, we describe the results of measurements
we performed on (almost fully) contained electromagnetic showers
with a PbWO$_4$ calorimeter. In that case, the asymmetry resulted from the integration over the full longitudinal shower profile, with the early part contributing most to the net asymmetry and the latest parts very
little, if anything. We measured for such showers an overall response asymmetry of 0.044 \cite{DREAM3show}, \ie $\sim 60\%$ of the asymmetry observed in the first $2-3 X_0$ reported here. 
\begin{figure}[htb]
\epsfysize=8cm
\centerline{\epsffile{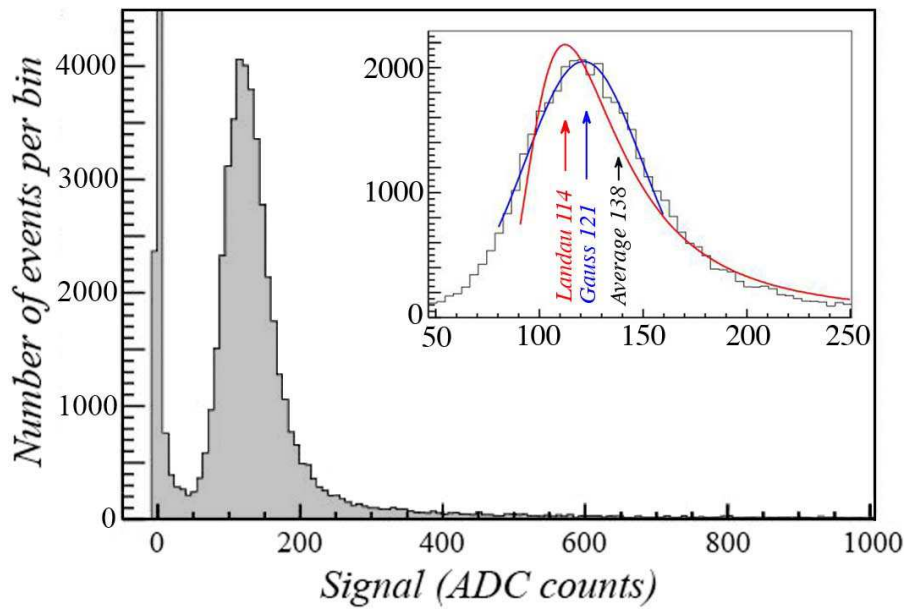}}
\caption{\small
Signal distribution for 150 GeV $\mu^+$ traversing the $2.5 X_0$ thick PbWO$_4$ crystal perpendicularly ($\theta = 0$). The insert shows the results of fits to the most probable signal region.
See text for details.}
\label{muonspec}
\end{figure}

The signals measured in these experiments were very small. According to EGS4 simulations, 10 GeV electrons deposited on average 320 MeV in the PbWO$_4$ crystal, when it was placed perpendicular to the beam line.
However, this was still one order of magnitude larger than the signals recorded for the muons.
Figure \ref{muonspec} shows a typical signal distribution, measured in one of the PMTs for $\theta = 0$.
Based on a comparison with the 10 GeV electron signals, this distribution gave a most probable energy deposit of 25 MeV.
\begin{figure}[htb]
\epsfysize=9cm
\centerline{\epsffile{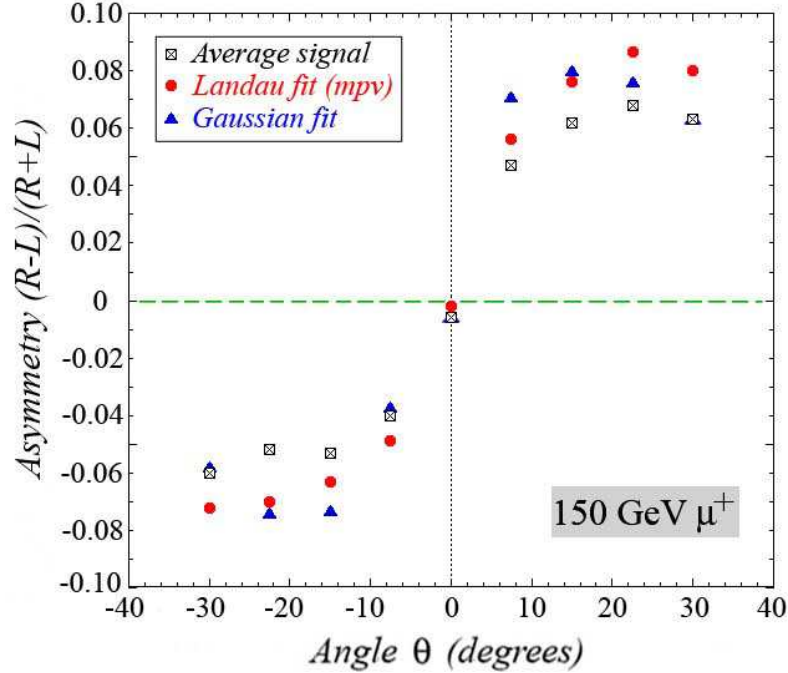}}
\caption{\small
Left-right response asymmetry measured for 150 GeV muons traversing a $2.5 X_0$ thick 
PbWO$_4$ crystal, as a function of the orientation of the crystal (the angle $\theta$). The asymmetry concerns the {\sl most probable} signal value derived from a Landau fit (the triangles) or a Gaussian fit (the closed circles), or the average signal value (the open squares).}
\label{Asymu}
\end{figure}

The most probable signal value was also used for the response asymmetry measurement for these particles. However,  the Landau fits did in general not reproduce the measured signal distributions very well.  This is because the beam spot was larger than the transverse size of the crystal. As a result, many beam particles missed the crystal (as evidenced by the large pedestal peak in Figure \ref{muonspec}), whereas others traversed it close to the edge, scattering out, leading to sub-mip signals.
For this reason, we also studied the left-right asymmetry using two other characteristics of the signal distribution: the average signal  and the result of a Gaussian fit around the most probable value (see insert Figure \ref{muonspec}).

The angular dependence of the left-right asymmetry for muons traversing the PbWO$_4$ crystal is shown in Figure \ref{Asymu}, for all three figures-of-merit derived from the signal distributions.
This curve shows the same general characteristics as the one measured for the electrons (Figure \ref{Asymel2}). The maximum asymmetry is measured for $\theta \approx \theta_C$ ($27^\circ$), in the PMT
oriented in the optimal direction for detecting the \v{C}erenkov light. The maximum asymmetry is $\sim 0.08$, which translates into a 15\% contribution  of \v{C}erenkov light to the total signals at that angle (Equation \ref{eq2}).
It also seems that the asymmetry is somewhat smaller when the average detector signal is used instead of the most probable value. This might indicate that \v{C}erenkov light produced in the radiative component of the energy lost by the muons is somewhat less directional than that produced by the ionizing component. This would be consistent with the observations of the asymmetry in electromagnetic showers discussed above.

Based on the signal distributions observed in these measurements and on the assumption that the
energy deposition by 10 GeV electrons is calculated correctly by GEANT4, we can determine the specific energy loss of muons in PbWO$_4$. The most probable energy loss is 11.6 MeV/cm, while the measured average energy lost by 150 GeV $\mu^+$ is 13.2 MeV/cm.
\begin{figure}[htb]
\epsfysize=7cm
\centerline{\epsffile{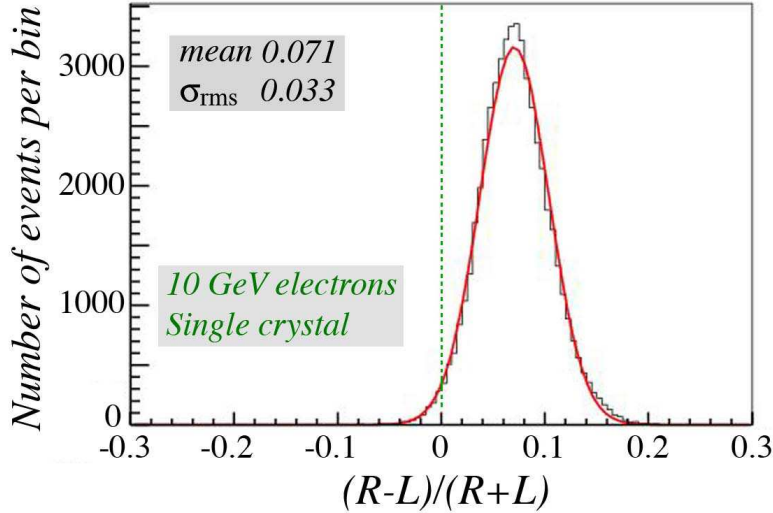}}
\caption{\small
Distribution of the left-right response asymmetry, measured for 10 GeV electrons traversing the crystal at  $\theta = 30^\circ$, together with the results of a Gaussian fit. }
\label{resol}
\end{figure}

We can also determine the ``light yield'' of the crystals and PMTs used in our studies, or rather the number of photoelectrons measured (with the PMTs chosen for these studies) per unit deposited energy.
Figure \ref{resol} shows the event-to-event distribution of the left-right asymmetry measured for
10 GeV electrons traversing the crystal at $\theta = 30^\circ$, together with the results of a Gaussian fit.
At $\theta = 0^\circ$, the width of the distribution increased from 0.0334 to 0.0384. 
If we assume that the width is dominated by statistical fluctuations in the number of photoelectrons, then a $\sigma$ of 0.0384 translates into an average number of photoelectrons of $\sim 340$ per PMT, or $\sim 1$ photoelectron per MeV deposited energy. At $\theta = 0^\circ$, the width of the $(R-L)/(R+L)$ distribution for the 150 GeV $\mu^+$ was 0.1274, which translates into an average signal of 31 photoelectrons per PMT.

The average signal from the 10 GeV electrons increased by 47\% when the crystal was rotated from 0$^\circ$ to 30$^\circ$.
This reflects the onset of the shower development, since a simple increase in path length ($\sim \cos^{-1}{\theta}$) would only lead to a 15\% increase. The measured width of the asymmetry distribution at 30$^\circ$ ($\sigma = 0.0334$) corresponds to the statistical fluctuations in 448 photoelectrons/PMT, an increase of only 32\% with respect to the 0$^\circ$ case. The fact that the decreasing width does not match the increased signal indicates that other factors (\eg temperature effects, response non-uniformities) did contribute to the measured width and that, therefore, the estimated light yield is in fact a lower limit.

The distribution shown in Figure \ref{resol}  illustrates one other important aspect of these experimental
data. All the measurement results reported in this paper concern {\sl  averages}. Both the observed left/right asymmetries, and also the angular dependence of the time structure discussed in the next subsection, concern the average characteristics of a large number of events, typically 100 000. The question arises how accurately one can determine the \v{C}erenkov content of the signal from one particular event on the basis of these characteristics. Figure \ref{resol} provides an answer to that question, for what concerns the left/right asymmetry.
It shows that, if for a particular 10 GeV electron traversing this crystal, an asymmetry is measured of 0.080, the
experimental uncertainty on this number is 0.033 (1 standard deviation). In other words, that particular signal contains 14.8$\pm$5.9\% \v{C}erenkov light (Equation \ref{eq2}). Of course, this error bar is strongly determined by photoelectron statistics, especially at these low energies ($\sim 0.4$ GeV).

We want to re-emphasize that all these results concerns one particular crystal, operated at room temperature.
No attempts were made to control the temperature, or to measure a temperature dependence of the observed effects.

\subsection{Time structure of the signals}

\begin{figure}[htb]
\epsfysize=8cm
\centerline{\epsffile{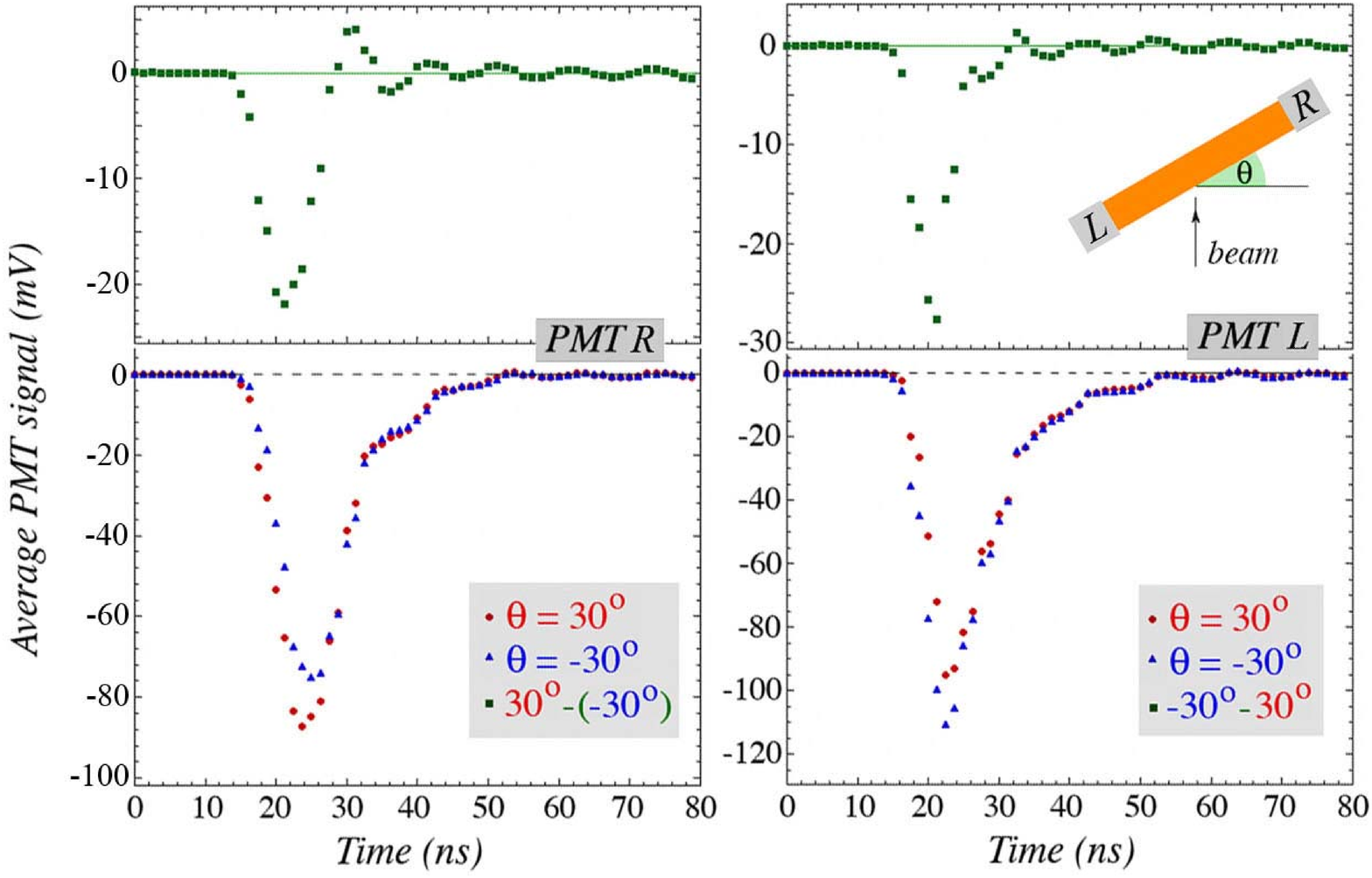}}
\caption{\small
 Average time structures of the signals measured in the left (L) and right (R) photomultiplier tubes that detect the light produced by 10 GeV electrons in a $2.2 X_0$ thick 
PbWO$_4$ crystal. The bottom plots show these signals for angles $\theta = \pm 30^\circ$, for PMTs R and L, respectively. The top plots show the difference between the two orientations, \ie the PMT's response function to a prompt \v{C}erenkov component in the signal.}
\label{Aeearly3}
\end{figure}

A second valuable tool for recognizing the contributions of \v{C}erenkov light to the calorimeter signals
is derived from the time structure of the events. This is illustrated in Figure \ref{Aeearly3}, which
shows the average time structure of the 10 GeV shower signals recorded with the same PMT at different angles  of incidence, namely $\theta =  30^\circ$ and $\theta =  -30^\circ$.
The diagrams on the left hand site of this figure concern PMT $R$.  \v{C}erenkov light produced by the traversing electron and by the particles produced in the early shower component are expected to be detected by this PMT when the crystal is oriented at $\theta =  30^\circ$, while very little, if anything, will reach this PMT at $\theta = -30^\circ$.   
The figure shows that the trailing edges of both time structures are almost completely identical.
This part of the time structure of the pulses is completely determined by the decay characteristics of the scintillation processes in the PbWO$_4$ crystals 
and should thus indeed be independent of the detector orientation.

However, there is a very significant difference in the leading edge of the pulses. The ones measured
for $\theta = 30^\circ$  exhibit a steeper rise than the ones for $\theta = -30^\circ$. The top graphs show the result of subtracting the latter pulse shape from the ``30$^\circ$'' one: The pulses recorded at $\theta = 30^\circ$ contain an additional ``prompt'' component of the type one would expect from
\v{C}erenkov light. 

The reverse situation is observed in the other PMT ($L$). Here, the prompt additional component is observed in the time structure of the pulses recorded when the crystal was oriented at $\theta =  -30^\circ$. Also in this PMT, the time structure beyond the amplitude of the signals was found to be independent of the crystal orientation. 

These results confirm the prompt nature of the additional light component observed in the left/right asymmetry measurements, and provide more evidence for the \v{C}erenkov mechanism being responsible for the observed phenomena.

The figure shows minor differences between the shapes of the prompt components observed in the
two PMTs. These are most likely due to differences in the characteristcs of these tubes. PMT $L$, which
operated at a slightly higher high voltage, responded to a $\delta$-function with a $\sigma_{\rm rms}$ of 2.3 ns, \vs 2.8 ns for PMT $R$. As a result, the response to the \v{C}erenkov component had a slightly different time structure in these PMTs.

The average pulse shapes shown in Figure \ref{Aeearly3} make it also possible to determine the contribution of \v{C}erenkov light to the crystal signals. 
In the case of the 10 GeV electrons, the additional component represented $\sim 12\%$
of the total signal in PMT $R$ and 13\% in PMT $L$. For comparison, we recall that the left/right asymmetry measurements led us to conclude that, at the \v{C}erenkov angle, these signals contained, on average, $\sim 13\%$ of \v{C}erenkov light\footnote{As before, these percentages represent the ratio of the \v{C}erenkov component and the total signal measured at the angle at which the contribution of this component is largest.}.

\begin{figure}[htb]
\epsfysize=11cm
\centerline{\epsffile{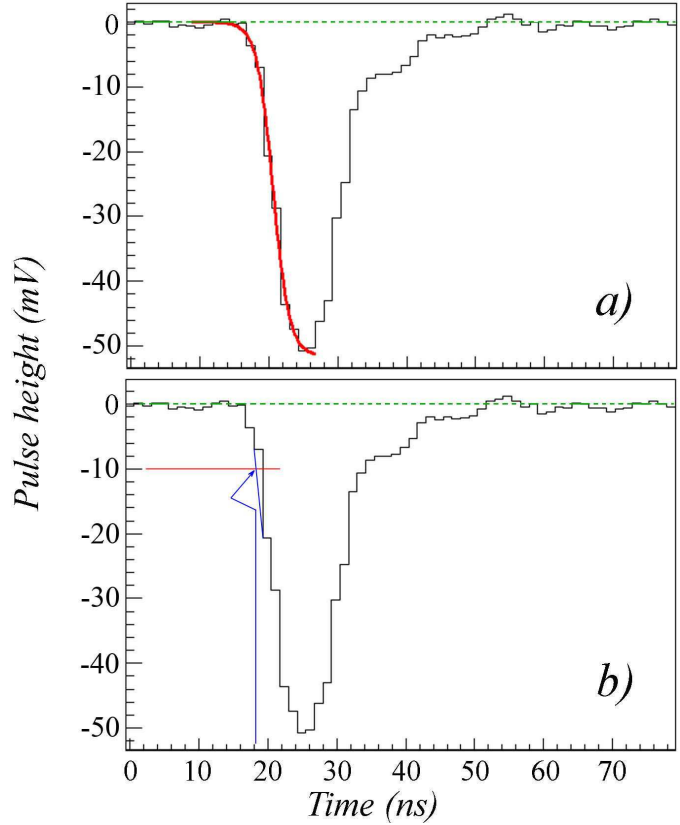}}
\caption{\small
The characteristics of the time structure of the signals are determined with two different methods. In method $a$, the leading edge is fitted to Equation \ref{eq3}, and the fitted parameters $t_L$ and $\tau_L$ determine the \v{C}erenkov content of the signal. In method $b$, the time at which the pulse height exceeds
a certain threshold level is used for this purpose. See text for details.}
\label{tmethod}
\end{figure}

We have also studied the angular dependence of the average pulse shape. Since the \v{C}erenkov contribution
manifests itself in the leading edge of the time structure, we have developed several methods to characterize the properties of that part of the pulse shape in a quantitative manner. Two of these methods are illustrated in Figure \ref{tmethod}.
In the first method (Figure \ref{tmethod}a), we used an appropriate function to describe the time structure. It turns out that
the leading edge of the pulse shape, $V(t)$, is well described by a function of the following type:
\begin{equation}
V(t) = |A|~ \Bigl[{1 \over {e^{(t-t_L)/\tau_L} + 1}} - 1\Bigr]
\label{eq3}
\end{equation}

The characteristics of the leading edge of the pulse shape are then determined by the values of the {\sl lead time} $t_L$ and the {\sl lead constant}  $\tau_L$, which are independent of the amplitude  $A$ of the signal. For example, an increase in the \v{C}erenkov content of the signal will manifest itself as a decrease in the value of 
$\tau_L$, since the leading edge is becoming steeper.

In the second method (Figure \ref{tmethod}b), we determined the precise time at which the pulse height exceeds a certain fixed threshold level, \eg -50 mV. An increase in the \v{C}erenkov content of the signal will
shift that point to an earlier moment.

\begin{figure}[htb]
\epsfysize=7cm
\centerline{\epsffile{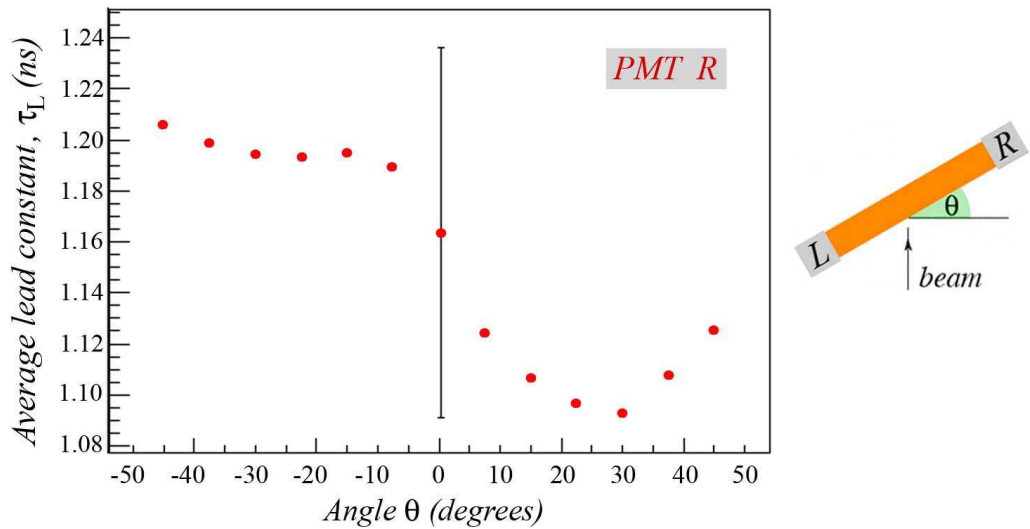}}
\caption{\small
Average lead constant, $\tau_L$ (see Equation \ref{eq3}), of the pulses recorded by PMT $R$, as a function of the orientation of the crystal, \ie the angle $\theta$. Data for 10 GeV electrons.}
\label{leadtime}
\end{figure}

\begin{figure}[htb]
\epsfysize=7cm
\centerline{\epsffile{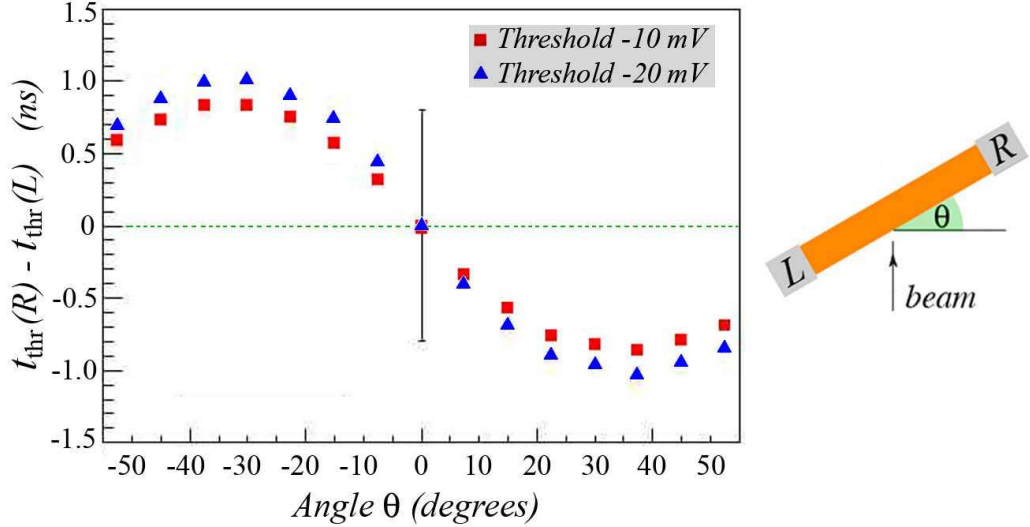}}
\caption{\small
Average difference between the times the two PMTs reading out the two sides of the crystal needed to reach a certain threshold level, as a function of the orientation of the crystal (\ie the angle $\theta$). Data for 150 GeV $\mu^+$, and two different threshold values.}
\label{mu_Scurve}
\end{figure}

Some results of these analyses are shown in Figures \ref{leadtime} and \ref{mu_Scurve}.
Figure \ref{leadtime} shows the value of the lead constant, $\tau_L$, measured for the 10 GeV electron signals from PMT $R$, as a function of the angle $\theta$. For negative values of $\theta$, \v{C}erenkov light produced
by the electrons was not detected by this PMT. The pulse shape is independent of $\theta$, with a $\tau_L$
value of 1.20 ns. However, when the crystal was rotated towards values of $\theta > 0$, \v{C}erenkov light produced
by the showering particles became a significant component of the signals measured by PMT $R$, and the leading edge of the pulse shape steepened ($\tau_L$ became smaller). This process continued until $\theta$ reached the \v{C}erenkov angle ($\sim 30^\circ$), at which point $\tau_L$ reached a minimum value of $\sim 1.09$ ns. For larger angles, the acceptance of \v{C}erenkov light decreased again and the leading edge became less steep, $\tau_L$ increased. The $\tau_L$ value measured for the signals from PMT $L$ shows a similar behavior:
It is constant for $\theta > 0$, decreases for $\theta < 0$, reaches a minimum value for $\theta = -30^\circ$ and increases again for larger angles.

Figure \ref{mu_Scurve} shows the results of an analysis of the threshold crossing time, for signals generated by 150 GeV $\mu^+$. The {\sl difference} between the crossing times of the signals recorded by PMTs $R$ and $L$ is given as a function of the angle $\theta$, for two different threshold levels. This difference is set to zero for $\theta =  0$. The figure shows that the difference is negative for $\theta > 0$.
As a result of the increasing contribution of \v{C}erenkov light to the signals from PMT $R$, the threshold was crossed earlier in this PMT, and therefore the plotted quantity is negative. When $\theta < 0$, \v{C}erenkov light contributed to the signals from PMT $L$, and the plotted quantity is positive. As in the case of the response asymmetry, the maximum difference is observed for angles near the \v{C}erenkov angle, $\sim 30^\circ$. The results are qualitatively not significantly different for the two different threshold levels, but they do  indicate a slightly larger effect for the higher threshold.

As in the case of the left/right asymmetry, the results shown in Figures \ref{leadtime} and \ref{mu_Scurve} concern
the average behavior observed for large numbers of events (100 000). The error bars in these figures indicate the
precision with which the parameter in question is determined for {\sl individual} events. The size of these error bars, which for these small signals is completely dominated by photoelectron statistics, is such that the value of
the lead constant ($\tau_L$) or the threshold crossing time does not provide statistically significant information about the (size of the) contribution of \v{C}erenkov light to the signal in question.

\section{Conclusions}

We have measured the contribution of \v{C}erenkov light to the signals from electrons and muons in lead tungstate crystals. In the chosen geometry, which was optimized for detecting this component, information about this contribution was obtained from 
the left/right response asymmetry  and from the time structure of the signals. For single particles traversing the calorimeter (muons), the maximum \v{C}erenkov contribution to the signals was measured to be 
$15 - 20\%$. The measurements for electron showers indicated somewhat lower values, because
of the contributions of isotropically distributed shower particles to the signals. This reduced the measured asymmetries in the response and time structure of the signals. This effect was measured
to increase in importance as the shower developed. The asymmetries measured for 10 GeV electrons in the first 2-3 radiation lengths were about three times larger than those measured at a depth of $7-10 X_0$, \ie just beyond the shower maximum.

\section*{Acknowledgments}

The studies reported in this paper were carried out with PbWO$_4$ crystals made available to us by the
PHOS group of the ALICE Collaboration. We sincerely thank Drs. Mikhail Ippolitov and Hans Muller 
for their help and generosity in this context.
We thank CERN for making particle beams of excellent quality available.
This study was carried out with financial support of the United States
Department of Energy, under contract DE-FG02-95ER40938.

\bibliographystyle{unsrt}

\end{document}